\documentclass{hcj}
\usepackage{subfigure}
\usepackage{multirow}
\usepackage{threeparttable}
\usepackage{color}

\newcommand{\blank}[1]{\hspace*{#1}\linebreak[0]}

\journalyear{2014}
\journalvolume{1}
\journalissue{2}
\journalpages{245-264}
\articledoi{10.15346/hc.v1i2.12}
\journalcopyright{\copyright{} 2014, Ponciano \& Brasileiro. CC-BY-3.0}
\title{Finding Volunteers' Engagement Profiles in \\ Human Computation for Citizen Science Projects}

\author{Lesandro Ponciano\affil{Universidade Federal de Campina Grande} \and
Francisco Brasileiro\affil{Universidade Federal de Campina Grande}}
\authorrunning{L. Ponciano and F. Brasileiro}

\begin{document}
 

\maketitle

\begin{abstract}

Human computation is a computing approach that draws upon human cognitive abilities to solve computational tasks for which there are so far no satisfactory fully automated solutions even when using the most advanced computing technologies available. Human computation for citizen science projects consists in designing systems that allow large crowds of volunteers to contribute to scientific research by executing human computation tasks. Examples of successful projects are Galaxy Zoo and FoldIt. A key feature of this kind of project is its capacity to engage volunteers. An important requirement for the proposal and evaluation of new engagement strategies is having a clear understanding of the typical engagement of the volunteers; however, even though several projects of this kind have already been completed, little is known about this issue. In this paper, we investigate the engagement pattern of the volunteers in their interactions in human computation for citizen science projects, how they differ among themselves in terms of engagement, and how those volunteer engagement features should be taken into account for establishing the engagement encouragement strategies that should be brought into play in a given project. To this end, we define four quantitative engagement metrics to measure different aspects of volunteer engagement, and use data mining algorithms to identify the different volunteer profiles in terms of the engagement metrics. Our study is based on data collected from two projects: Galaxy Zoo and The Milky Way Project. The results show that the volunteers in such projects can be grouped into five distinct engagement profiles that we label as follows: hardworking, spasmodic, persistent, lasting, and moderate. The analysis of these profiles provides a deeper understanding of the nature of volunteers' engagement in human computation for citizen science projects.\\

\noindent {\bf keywords}: citizen science, human computation, engagement, participation, retention
\end{abstract}

\section{Introduction}

Human computation is a computing approach based on harnessing human cognitive abilities to solve computational tasks for which there are so far no satisfactory fully automated solutions even when using the most advanced computing technologies currently available~\cite{Quinn2011}. Examples of such tasks may be found in the areas of natural language processing, image understanding, and creativity. They have been shown to be often in scientific applications related to disciplines such as biology, linguistics, and astronomy~\cite{Wiggins2012,lintott:2013}. As a result, it has become common among scientists to start projects to recruit ordinary people for executing human computation tasks, which we call {\it human computation for citizen science projects}. Citizen science can be broadly defined as a partnership between scientists and ordinary people willing to contribute to an authentic scientific research effort~\cite{Cohn2008,Dickinson2012,lintott:2013}. A large range of activities can be carried out by ordinary people in citizen science~\cite{Michael2007,Cohn2008,Wiggins2012}. Those activities may require only some simple abilities, such as data collecting and reporting, or more complex cognitive abilities such as data aggregation and classification. In human computation for citizen science projects, participants contribute by executing tasks that require cognitive abilities. Examples of projects with such feature are Galaxy Zoo~\cite{Lintott:2008} and FoldIt~\cite{cooper:2010}.

The contribution behaviour of people taking part in this type of project can be examined in the light of two different research approaches centered on the notions of {\it voluntarism}~\cite{clary:1998,wilson:2000} and {\it human engagement}~\cite{Brien:2008,simpson:2009,Lehmann:2012}. Voluntarism literature usually distinguishes between two different types of contribution behaviour: helping activity behaviour and volunteerism behaviour~\cite{clary:1998,wilson:2000}. Helping activity behaviour designates a form of \emph{sporadic} participation in which the individual is faced with an unexpected request to help someone to do something. Volunteerism behaviour, on the other hand, concerns to a kind of \emph{planned} behaviour. Volunteers are usually actively seeking out opportunities to help others. They typically commit themselves to an ongoing relationship at considerable personal cost in terms of dedicated time or cognitive effort. Drawing this distinction between helping activity and voluntarism seems to us to be important also in the context of human computation for citizen science projects. A recent characterization of the behaviour of volunteers in such projects brings to light the existence of two main groups of participants: transient and regular~\cite{Ponciano:CiSE:2014}. Transient participants exhibit a helping behaviour, whereas the behaviour of regular participants fits into the definition of volunteerism. Not surprisingly, volunteers typically constitute a minority among the participants, and execute the largest part of tasks in the project. Thus, a key feature for the success of a human computation for citizen science project is the capacity to foster such kind of sustained contribution behaviour.

Fostering sustained contribution behaviour is an issue that has been widely addressed in human engagement studies. Current literature on human engagement focuses on the human behaviour when individuals are self-investing personal resources such as time, physical energy, and cognitive power~\cite{bakker:2008,Brien:2008,simpson:2009,Lehmann:2012,McCay-Peet:2012}. Studies in this area usually focus on both qualitative and quantitative dimensions of engagement by (i) analysing the psychological factors behind engagement/disengagement such as motivation, satisfaction, and frustration; and (ii) measuring the level of engagement quantitatively in terms of the degree of contribution and the duration of the contribution.

Several studies have been devoted to the understanding of psychological factors of volunteer engagement in human computation for citizen science projects~\cite{raddick:2010,rotman:2012,jennett:2014,Nov2014}, while few studies have focused on quantitatively estimation of the level of engagement of the volunteers~\cite{Ponciano:CiSE:2014}. The lack of studies with this perspective is an important constraint because a fundamental requirement for proposing and evaluating new engagement strategies is having a clear understanding of how volunteers typically behave in such situations. This study aims at filling this gap by providing a quantitative analysis of the nature of engagement of volunteers by using log data related to their execution of tasks. Three research questions are addressed in this study: $1)$ how engaged the volunteers are during their interaction with the project; $2)$ what similarities and differences they exhibit among themselves in terms of engagement; and $3)$ how the engagement characteristics of the volunteers can be exploited for establishing the engagement strategies to be implemented in a given project.

In order to answer these questions, we go through existing human engagement studies and, based on the concepts and theories put forward, we propose the following four metrics to measure the level of engagement of each volunteer: activity ratio, relative activity duration, daily devoted time, and variation in periodicity. Activity ratio is a measure of the return rate of the volunteer to the project during the period that he/she stays contributing to it. Daily devoted time is a measure of the length of the daily engagement. Relative activity duration, in turn, is a measure of the duration of the volunteer's long-term engagement. Finally, variation in periodicity informs us about the deviation in the periodicity with which the volunteer executes tasks in the project. By using hierarchical and k-means algorithms, we cluster the volunteers according to the values of their engagement metrics in order to find out the different engagement profiles that arise from their natural behaviour within the project.

We analyse volunteer engagement profiles according to the data collected from two popular projects hosted at the Zooniverse platform: Galaxy Zoo and The Milky Way Project. These projects ran for almost 2 years between 2010 and 2012 and involved more than one billion executed tasks and thousands of participants, which turns them into valuable sources for the analysis of a wide range of engagement aspects of the volunteers. In both projects, we found 5 different clusters of volunteers based on visual inspection and statistical measures. Each cluster stands for a distinct engagement profile brought for by the behaviour shown by the volunteers during their participation in the projects. The distinct engagement profiles brought to light in this way are labelled as: hardworking, spasmodic, persistent, lasting, and moderate.

Hardworking engagement is characterised by larger activity ratio, low variation in periodicity and shorter relative activity duration. Volunteers who exhibit this type of engagement profile typically work hard and regularly when arriving at the project, but may leave the project quickly. Spasmodic engagement is distinguished by a relatively high activity ratio and moderate variation in periodicity. Volunteers who exhibit this engagement profile provide an intense contribution, at a short period of time and with irregular periodicity within this period. Persistent engagement, in turn, is characterised by a larger activity duration and low activity ratio. Volunteers who exhibit a persistent engagement profile remain in the project for a long period of time but contribute only a few days within this time period. Lasting engagement, in turn, is characterised by  an engagement pattern similar to persistent engagement, with the difference that volunteers exhibit here a much shorter activity duration. Finally, moderate volunteers have intermediate scores in all categories of engagement metrics.

Regarding the distribution of the volunteers per profile, the highest percentage of volunteers ($30\%$ in The Milky Way Project and $31\%$ in Galaxy Zoo) exhibits a moderate engagement profile, while few volunteers ($13\%$ in The Milky Way Project and $16\%$ in Galaxy Zoo) show persistent engagement. Given the total amount of human effort time required to execute all the tasks in the project, the aggregate time devoted by volunteers who exhibit a persistent engagement profile accounts for $40\%$ of total time in The Milky Way Project and $46\%$ in Galaxy Zoo; this is the volunteer profile that stands for the largest contribution.

The method we propose to measure the engagement of volunteers and set up engagement profiles has been shown to be satisfactory in bringing to light the main similarities and differences among the volunteers. The fact that the results thus obtained are consistent throughout different projects strengthens the thesis that engagement profiles can arise in various other projects. Several other discussions can be drawn from our analysis. For example, the engagement profiles enable the development of new recruitment strategies to attract volunteers with a desired engagement profile as well as the design of personalised engagement strategies that focuses on improving specific engagement metrics. Finally, our results call for further theoretical and qualitative studies that investigate the motivation of volunteers in the light of the distinct engagement profiles they may exhibit. The combination of a quantitative analysis of volunteer engagement and the psychological factors established in qualitative studies will advance our comprehension about the engagement patterns of volunteers in human computation and citizen science.

In this study we put forward three main contributions. First, we propose four metrics to measure the level of engagement of volunteers with regard to both the duration of the period of engagement with the project and the degree of engagement during this period. Furthermore, we provide a deeper quantitative assessment of volunteer engagement profiles derived from two popular human computation for citizen science projects. To the best of our knowledge, this is the first study assessing natural engagement profiles in volunteer task execution behaviour in this type of project. Finally, this study allows us to go beyond previous studies by covering a larger number of volunteers and bringing forth engagement aspects which have so far not been identified in studies focusing on qualitative methodologies.

The rest of this work is organised as follows. We provide first a background of human engagement studies and discuss relevant previous work. Next we describe our method to measure the volunteer engagement and identify engagement profiles. Finally, we present an analysis of volunteer engagement in Galaxy Zoo and The Milky Way Project.

\section{Background and Related Work}

This study builds on a broad set of studies covering volunteer engagement, human computation and citizen science projects. In this section, we first provide a background to the subject of human engagement. Thereafter, we discuss the related work.

\subsection{What is engagement and how to approach it}

The subject of human engagement has been studied within a variety of disciplines, such as education~\cite{meece1988}, management science~\cite{simpson:2009} and computer science~\cite{Brien:2008}. Some studies make an attempt to conceptualize the term {\it engagement} in an interdisciplinary perspective~\cite{gonzalez:2006,bakker:2008,Brien:2008,simpson:2009,Lehmann:2012,McCay-Peet:2012}. A consensus that emerges from these studies is that engagement means to participate in any enterprise by self-investing personal resources, such as time, physical energy, and cognitive power.

\citet{Brien:2008} provide a conceptual framework to study human engagement with technology. This framework establishes that the entire process of engagement is comprised of four stages: point of engagement, period of sustained engagement, disengagement and reengagement. The {\it point of engagement} is the time at which the human perform the first action in the system. The {\it period of sustained engagement} is the continuous period of time in which he/she keeps on performing actions in the system. {\it Disengagement} occurs when the period of sustained engagement ends. Finally, {\it reengagement} denotes new engagement cycles composed of point the three first stages. Studies of such process involve at least four dimensions: type of engagement, psychological factors of engagement, duration of engagement, and degree of engagement.

The {\it type of engagement} is defined by the kind of personal resources and skills that humans invest in performing an activity. Examples of types of engagement are social engagement~\cite{Porges:2003} and cognitive engagement~\cite{corno1983}. Social engagement refers to actions that require humans to interact with others. It is widely studied in areas such as online social networks and communities~\cite{Preece:2000,millen:2002}. Cognitive engagement refers to actions that require mainly human cognitive effort. It has been widely addressed in educational psychology and work engagement~\cite{meece1988,simpson:2009}.

The {\it psychological factors of engagement} are related to the motives leading to a point of engagement, disengagement and reengagement, such as motivation, satisfaction, perceived control, and frustration. Studies have proposed and/or instantiated various theories in order to construct a framework of theories that explain the psychological factors behind human engagement~\cite{gonzalez:2006,Brien:2008}. These theories include the self-determination theory~\cite{Deci2000} and the self-efficacy theory~\cite{bandura1977}. The self-determination theory establishes that human motivation can be broadly divided into intrinsic motivations, associated with inner personal reward, and extrinsic motivations, associated with earning an external reward or avoiding a punishment. The self-efficacy theory, in turn, advances the idea that perceived human efficacy determines if an individual will initiate an activity, how much effort will be expended, and how long the activity will be sustained.

The {\it duration of engagement} measures the duration of the period of sustained engagement, sometimes called retention. It expresses how long a human keeps on to the system. It is short-term engagement when it occurs during a relatively short period of time (e.g. minutes or hours), and long-term engagement when it lasts a long period of time (e.g. months or years). In short-term engagement, the point of engagement is the point in time at which the individual performs the first action within the system, the period of engagement is the time span under which he/she keeps interacting with the system in a continuous working session, and the point of disengagement is the point in time at which the working session ends. In long-term engagement, the point of engagement is the point in time at which the individual performs the first action within the system, the period of engagement refers to the number of days under which she/he keeps on interacting with the system, and the point of disengagement refers to the day when he/she leaves the system. Thus, long-term engagement may consist of several short-term engagement cycles.

Finally, the {\it degree of engagement} is a quantitative measure of the degree of participation during the period of sustained engagement. It can also be viewed as a measure of the amount of resources invested by humans in participating in the system. Measuring the degree of engagement has proven a challenging task. Some studies use surveys to collect information about how humans perceive their level of engagement and hence estimate their degree of engagement (e.g.,~\citet{Brien:2010,McCay-Peet:2012}). Other studies use behavioural data stored in logs of the system to measure the degree of engagement (e.g.~\citet{Lehmann:2012}).

\subsection{Related work}

The dimensions of engagement presented in the last section are helpful to framing the previous studies in engagement. There is an extensive body of work dealing with engagement in technology-mediated social participation systems~\cite{Kraut:2010} such as wiki-based systems~\cite{butler2002,Bryant:2005,Butler:2008,Schroer:2009,Preece2009,niederer2010,Liu:2011,Welser:2011,Zhu:2012}, open source software projects~\cite{Hertel2003,niederer2010}, and human computation for citizen science projects~\cite{raddick:2010,rotman:2012,lopez:2012,mao:2013,jennett:2014}.

Wiki-based systems such as Wikipedia provide means that allow participants to engage in a broad range of activities, such as the insertion of a sentence in an article, modification of an existing reference, reverting an article to a former version etc~\cite{Butler:2008,Liu:2011,Welser:2011}. Participants assume different roles in the system when some of them focus on performing a single type of activity, and others focus on performing other types of activities~\cite{Butler:2008,niederer2010,Liu:2011}. Such roles characterise different types of engagement in the system. The motivation of the participants and their perception of their own roles usually change as they become more active in the system~\cite{Bryant:2005,Burke:2008,Schroer:2009,Preece2009}. Since such systems provide a collaborative environment, the behaviour of some of the participants may also affect the behaviour of others~\cite{butler2002,Zhu:2012}.

Studies on open source software (OSS) projects, in turn, have focused on understanding the psychological factors that lead participants to engage in OSS projects, and the kind of rewards they expect~\cite{Hertel2003,Roberts:2006}. For example, ~\citet{Hertel2003} show that psychological factors appeared to be similar to those behind voluntary action within social movements such as the civil rights, labour, and peace movements. Studies on Apache projects suggest that there are also interrelationships between motivation and degree of engagement~\cite{Roberts:2006}. Extrinsic motivation, such as monetary and status within the system, leads to above average contribution levels, while intrinsic motivations do not significantly impact average contribution levels.

Differently from Wiki-based systems, in which there is a diversity of types of engagement, the role played by volunteers in human computation for citizen science projects is mainly the execution of well defined human computation tasks, although some projects allow volunteers to carry out social engagement activities, for instance interacting in forums~\cite{Fortson:2011,luczak2014}. In such projects, as in the case of studies in wiki-based systems and OSS projects, the psychological factor is the dimension of engagement that has received most attention~\cite{raddick:2010,rotman:2012,jennett:2014,Nov2014}.

\citet{raddick:2010} analyse the motivations of volunteers in the Galaxy Zoo project. It is shown that, among $12$ categories of motivations mentioned by the volunteers, the most mentioned category is interest in astronomy, which is the theme of the project. \citet{rotman:2012} and \citet{Rotman:2014} show that the motivation of volunteers changes dynamically throughout the period of their contribution to the projects. \citet{jennett:2014} analyse factors that led volunteers to dabble and/or drop-out in the Old Weather project. The analysis shows that this kind of volunteers are less motivated, though they care about the project and the quality of the work they perform. Thus, projects should be designed to encourage both dabbling and commitment. \citet{Nov2014} analyses motivation factors that affect the quality and the quantity of contributions to citizen science projects.

In general, these studies clarify several aspects of {\it why} volunteer engages in human computation for citizen science projects. However, little progress has been made in terms of understanding {\it how} to measure volunteer engagement and to uncover natural patterns in which the engagement occurs. This fact constitutes an important shortcoming because a key feature of this kind of project is its capacity to engage volunteers. A clear understanding of how volunteers typically engage with such kinds of projects is fundamental for proposing and evaluating new strategies to encourage engagement.

\section{Finding Engagement Profiles}

In this section, we first present the metrics proposed to measure the degree of engagement and the duration of engagement of volunteers. Then, we present a strategy to cluster volunteer based on the values of these metrics for the volunteers. This clustering allows the identification of profiles of volunteers exhibiting similar engagement patterns.

\subsection{Measuring engagement}

We characterise volunteers according to how they score in different engagement metrics. Engagement metrics are measures of volunteer interaction and involvement with the project. The engagement metrics proposed in this section are based on the conceptual framework proposed by~\citet{Brien:2008}. By using this framework, we analyse the engagement over time of volunteers taking into account their points of engagement, periods of sustained engagement, disengagements and reengagements.

\begin{figure}[htb]
\centering
\includegraphics[width=1\linewidth]{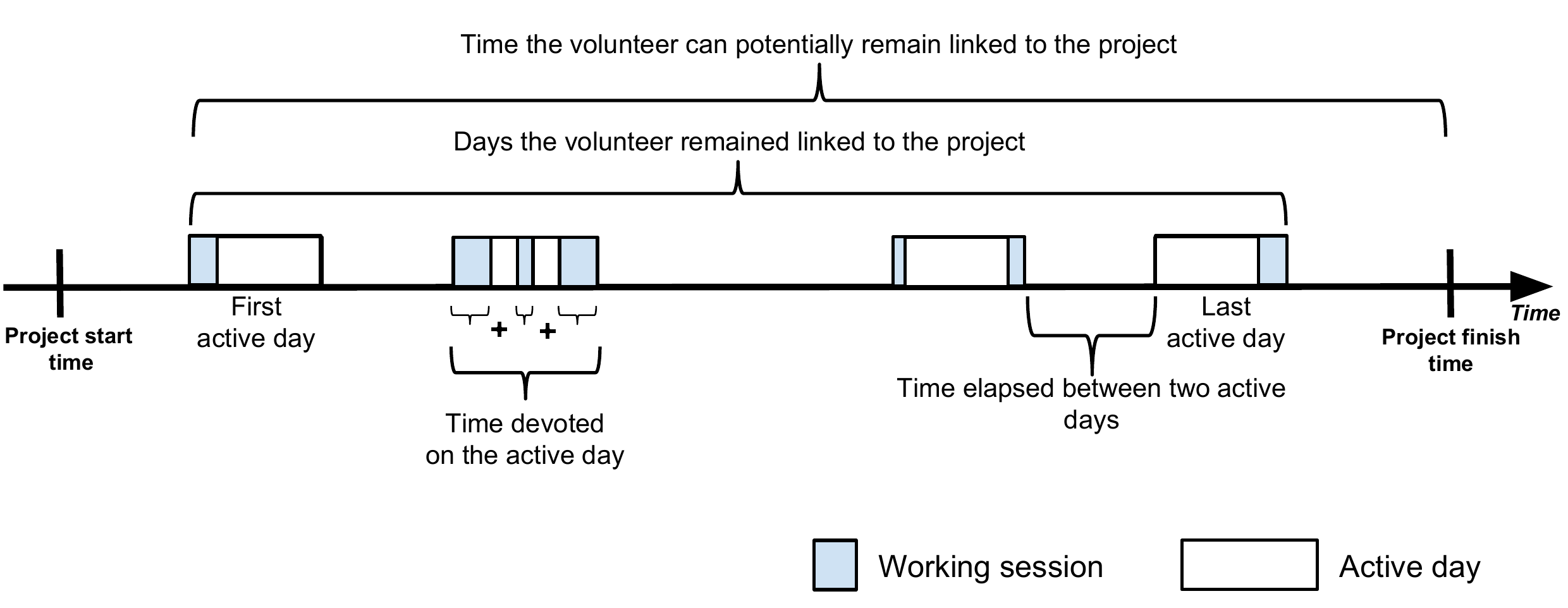}
\caption{Structure of the time line of a volunteer in a project, highlighting the active days and working sessions on the active days.}
\label{fig:volunteerTimeLine}
\end{figure}

Figure~\ref{fig:volunteerTimeLine} shows the structure of the time line of a volunteer during participation in a project. This figure shows five concepts used in the calculations of our metrics: the time the volunteer could potentially remain linked to the project, days the volunteer remain linked to the project, the active days, the time devoted on an active day, and the number of days elapsed between two active days. Our metrics are designed to measure the engagement of participants that exhibit an ongoing contribution and have contributed in at least two different days. By doing so, we focus on participants that are more likely to fit into the voluntarism definition~\cite{clary:1998,wilson:2000}.

The time a volunteer $i$ can potentially remain linked to the project is the number of days elapsed between the day in which the volunteer joined the project and the day in which the project is concluded. It is denoted by $w_i$ days. An {\it active day} of a volunteer $i$ is a day on which this volunteer is active in the project. We consider that a volunteer is active on a particular day if he/she executes at least one task during that day. We define $A_i$ as the sequence of dates in which the volunteer $i$ is active. The {\it time devoted on a specific active day} is the sum of the time duration of the contribution sessions of the volunteer on that active day. Contribution sessions are continuous short periods of time during which the volunteer keeps executing tasks. We define $D_i$ as the multiset of the amount of time the volunteer $i$ devotes to the project on each active day. The {\it time elapsed between two active days} is the number of days it took to the volunteer to return to the project since the latest active day. We define $B_i$ as the multiset of the number of days elapsed between every two sequential active days. Considering $w_i$, $A_i$, $D_i$ and $B_i$, we can derive metrics to measure the degree and the duration of engagement of each volunteer.

We define two metrics of degree of engagement: activity ratio and daily devoted time. {\bf Activity ratio ($a_i$)} is the proportion of days on which the volunteer was active in relation to the total of days he/she remained linked to the project. It can be computed as $a_i = \frac{|A_i|}{(Max(A_i) - Min(A_i))+1}$, $a\in(0,1]$. The closer to 1, the more assiduous the volunteer is during the time he/she remained linked to the project. {\bf Daily devoted time ($d_i$)} is the averaged hours the volunteer remain executing tasks on each day he/she is active. It can be computed as $d_i = avg(D_i)$, $d\in(0,24]$. The higher the average, the longer the time the volunteer devotes to the project executing tasks on the days he/she is active. Note that, because the human computation projects usually consist of different time-consuming tasks, the time devoted by the volunteers executing tasks is a better measure of their degree of engagement than the number of tasks they execute~\cite{geiger:2013,Ponciano:CiSE:2014}.

We also define two metrics to assess the duration of engagement: relative activity duration and variation in periodicity. {\bf Relative activity duration ($r_i$)} is the ratio of days during which a volunteer $i$ remains linked to the project in relation to the total number of days elapsed since the volunteer joined the project until the project is over ($w_i$). It is defined as $r_i = \frac{(Max(A_i) - Min(A_i))+1}{w_i}$, $r \in (0,1]$. When $r_i = 1$, the volunteer remains linked to project since she/he came to the project until the project is completed. The closer to $1$, the more persistent is the participation of the volunteer in the project. {\bf Variation in periodicity ($v_i$)} is the standard deviation of the times elapsed between each pair of sequential active days. It is computed as $v_i = sd(B_i)$. When $v_i = 0$, the volunteer exhibits a constant elapsed time between each pair of sequential active days; this indicates that he/she comes back to the project with perfect periodicity. On the contrary, the larger $v_i$, the larger the deviation in the periodicity in which the volunteer comes back to the project to perform more tasks.

The above engagement metrics fit well into our objective of analysing the degree of engagement and the duration of engagement of the volunteers. Activity ratio allows us to analyse the return rate of each volunteer to the project during the period that he/she stays contributing. Daily devoted time gives us a view of the length of the daily engagement, which is related to the duration of the short-term engagement. Relative activity duration allows us to analyse the duration of long-term engagement weighted by the duration of the period in which the volunteer can potentially remain linked to the project. Finally, variation in periodicity informs us about the periodicity of return during the long-term engagement.

\subsection{Clustering volunteers according to engagement metrics}

We use clustering algorithms to find out groups of volunteers who exhibit similar values for the engagement metrics. The input to clustering algorithms is a matrix $|I|\times4$ in which each row stands for a volunteer $i\in I$ and each column is an engagement metric, i.e. $a$, $d$, $r$, and $v$. As the results of clustering depend on the relative values of the parameters being clustered, a normalisation of the parameters prior to clustering would be desirable~\cite{jain2008}. We use range normalisation to scale the values of the engagement metrics in the interval $[0,1]$. The scaling formula is $x_i = \frac{x_i-x_{min}}{x_{max}-x_{min}}$, where $x$ denotes the engagement metric and $i$ the volunteer.

To identify the suitable number of clusters, we first run a hierarchical clustering algorithm and observe its dendrogram, which yields a suitable interval to test the number of clusters. Next we run k-means, varying the number of clusters ($k$) in the suggested interval and using as initial centroids the centres identified in the hierarchical clustering, which usually reduces the impact of noise and requires less iteration time~\cite{Lu2008}. We select thereafter a suitable $k$ and evaluate the quality of the clustering by computing the within-group sum of squares~\cite{anderberg1973} and Average Silhouette width~\cite{rousseeuw1987silhouettes}.

Within-group sum of squares measures the differences between the volunteers and the centre of the group to which they belong. The lower the within-group sum of squares, the better the clustering. It indicates that volunteers clustered in the same group exhibit similar values for the engagement metrics and that the centre of the group represents the group adequately. Average Silhouette width, in turn, measures how well separated and cohesive the groups are. This statistics ranges from $-1$, indicating a very poor clustering, to $1$, indicating an excellent clustering. \citet{Struyf1997} propose the following subjective interpretation of the silhouette statistics: between $0.71$ and $1.00$, a strong structure has been found; between $0.51$ and $0.70$, a reasonable structure has been found; between $0.26$ and $0.50$, the structure is weak and could be artificial, and hence it is recommended that additional methods of analysis are tried out; less than or equal to $0.25$, no substantial structure has been found. In this study, a silhouette statistics larger than or equal to $0.51$ indicates a reasonable partition of the different patterns of engagement exhibited by the volunteers.

\section{Engagement Profiles in Galaxy Zoo and The Milky Way Project}

In this section we use the proposed method to analyse the engagement of volunteers in two projects: Galaxy Zoo and The Milky Way Project. We first introduce these projects and detail the data set collected from them. Then, we present the results on the quality of clustering in these data sets and the discovered engagement profiles. Finally, we discuss the results and their implications.

\subsection{Datasets}

The data used in this study was collected from two human computation for citizen science projects: Galaxy Zoo Hubble and The Milky Way Project. Both projects were developed and deployed in the Zooniverse (zooniverse.org) citizen science platform.

The original Galaxy Zoo~\cite{Lintott:2008} was launched in July 2007, but has been thereafter redesigned and relaunched several times. In this project, participants were asked to answer a series of simple questions about the morphology of galaxies. Each classifying volunteer on Galaxy Zoo is presented with a galaxy image captured by either the Sloan Digital Sky Survey (SDSS) or the Hubble Space Telescope. A decision tree of questions is presented with the answer to each question being represented by a fairly simple icon. The task is straightforward and no specialist knowledge is required. In this paper, we used data of the third iteration of Galaxy Zoo: Galaxy Zoo Hubble. It was launched in April 2010 and ran until September 2012. It consisted of $9,667,586$ tasks executed by $86,413$ participants. In The Milky Way Project~\cite{simpson:2012}, participants are asked to draw ellipses onto the image to mark the locations of bubbles. A short online tutorial shows how to use the tool, and examples of prominent bubbles are given. As a secondary task, users can also mark rectangular areas of interest, which can be labelled as small bubbles, green knots, dark nebulae, star clusters, galaxies, fuzzy red objects or ``other''. Users can add as many annotations as they wish before submitting the image, at which point they are given another image for annotation. We used data of The Milky Way Project launched in December 2010 and ran until September 2012. It consisted of $643,468$ tasks executed by $23,889$ participants.

Each entry in the data set refers to one task execution. Each task execution is described by project\_id, task\_id, user\_id, datetime. The {\it project\_id} field is the name of the project. The {\it task\_id} field is a unique task identifier in the project. The {\it user\_id} field is a unique volunteer identifier in the project. Finally, the {\it datetime} field indicates the date and time when the task was executed. To form volunteers' working sessions, we use the threshold-based methodology~\cite{geiger:2013,mehrzadi:2012,Ponciano:CiSE:2014}. Following this methodology, we compute the interval of time elapsed between every two sequential task executions for each volunteer. Given these intervals, we use the method proposed by~\citet{mehrzadi:2012} to identify for each volunteer a threshold that distinguishes short intervals from long intervals. Hence, whenever the interval between the execution of two tasks is not larger than the threshold, the two tasks are assumed to have been executed in the same working session; otherwise, the tasks are assumed to have been executed in two different and consecutive working sessions. For more details about this methodology, see~\citet{mehrzadi:2012}.

In both projects, participants are considered volunteers only if they have been engaged in at least two days of activity. Only volunteers  who arrived before the last quarter of the total duration time of the project were considered in the analyses, i.e. the first 502 days of The Milky Way Project and the first 630 days of the Galaxy Zoo project. As Table~\ref{tab:descriptiveStatistics} shows, the final dataset consists of $23,547$ volunteers for the Galaxy Zoo and $6,093$ volunteers for The Milky Way Project, whereas 2485 volunteers contributed to both projects. As shown by the descriptive statistics in this table, in both projects the volunteers differ among themselves significantly in terms of all the engagement metrics, all of which are significantly non-normal (Kolmogorov-Smirnov normality tests showing p-value $< 0.05$). The variations in the engagement metrics of the volunteers do not point out at any form of anomalous behaviour among the volunteers, which can thus be considered as natural throughout.

\begin{table}
\footnotesize
\caption{Descriptive statistics of engagement metrics of volunteers in the studied datasets}
\label{tab:descriptiveStatistics}
\centering
\begin{tabular}{l | r | r }
            \hline
            & \multicolumn{1}{|c}{\bf The Milky Way Project} & \multicolumn{1}{|c}{\bf Galaxy Zoo}\\ \hline
            \#Volunteers & 6,093 & 23,547\\
            Activity ratio & $mean=0.40$, $sd=0.40$&$mean=0.33$, $sd=0.38$\\
            Daily devoted time & $mean=0.44$, $sd=0.54$& $mean=0.32$, $sd=0.40$\\
            Relative activity duration & $mean=0.20$, $sd=0.30$& $mean=0.23$, $sd=0.29$ \\
            Variation in periodicity &$mean=18.27$, $sd=43.31$&$mean=25.23$, $sd=49.16$\\\hline
\end{tabular}
\end{table}

\subsection{Clustering}

\begin{figure}[htb]
\vspace{-30pt}
\centering
\subfigure[Within-groups sum of squares]{
\includegraphics[width=.45\linewidth]{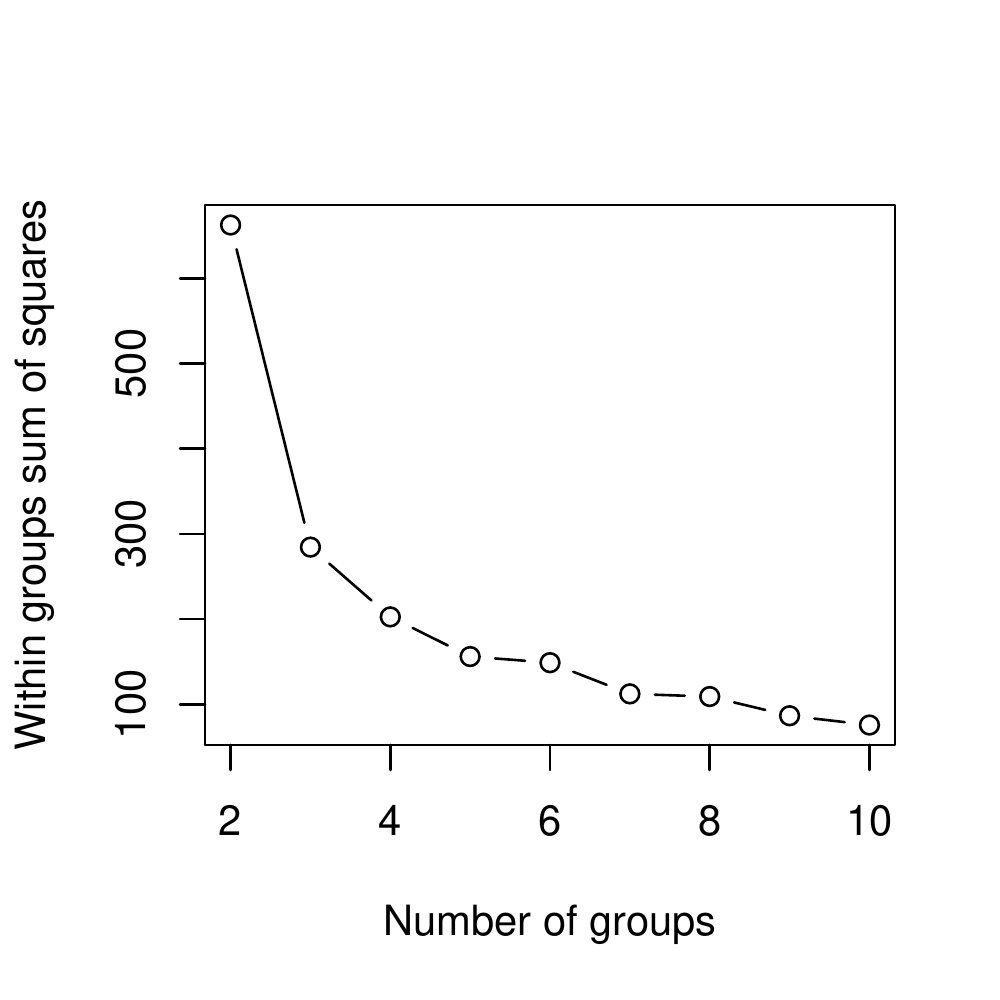}
\label{subfig:mw-withinGroupsSumOfSquares}
}
\hspace{-10pt}
\subfigure[Average Silhouette statistic]{
\includegraphics[width=.45\linewidth]{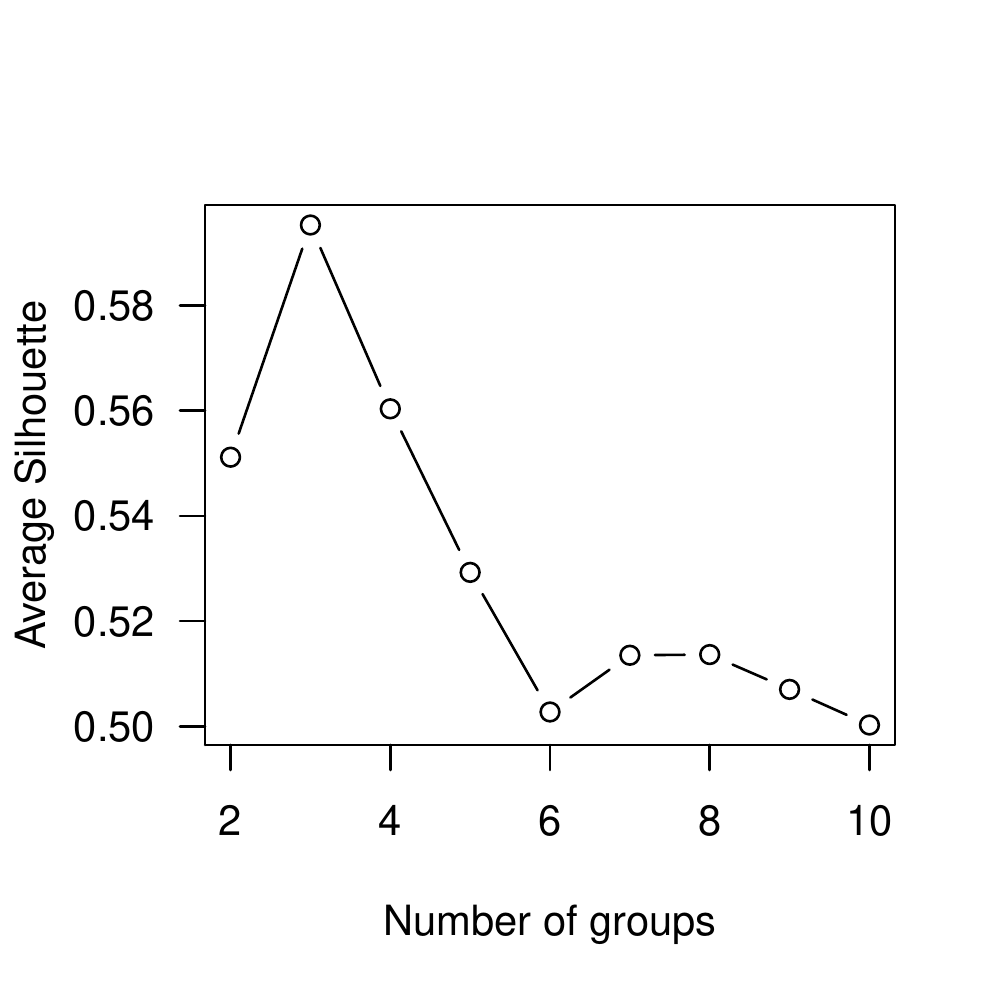}
\label{subfig:mw-averageSilhouette}
}
\vspace{-3pt}
\caption{Analysis of k-means clustering in The Milky Way Project. Within-groups sum of squares and average Silhouette statistic as the number of groups (k) is varied.}
\label{fig:mw-clusterEvaluation}
\vspace{-5pt}
\end{figure}

\begin{figure}[htb]
\vspace{-30pt}
\centering
\subfigure[Within-groups sum of squares]{
\includegraphics[width=.45\linewidth]{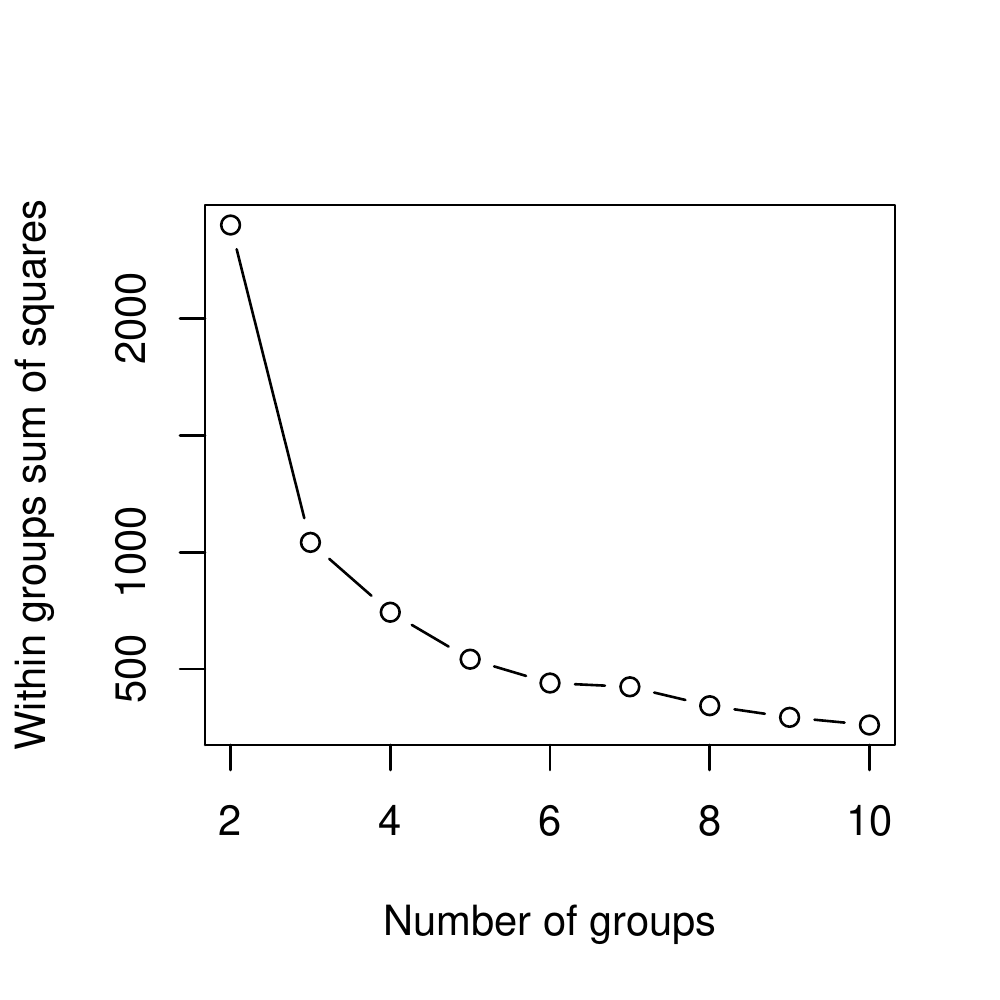}
\label{subfig:gz-withinGroupsSumOfSquares}
}
\hspace{-10pt}
\subfigure[Average Silhouette statistic]{
\includegraphics[width=.45\linewidth]{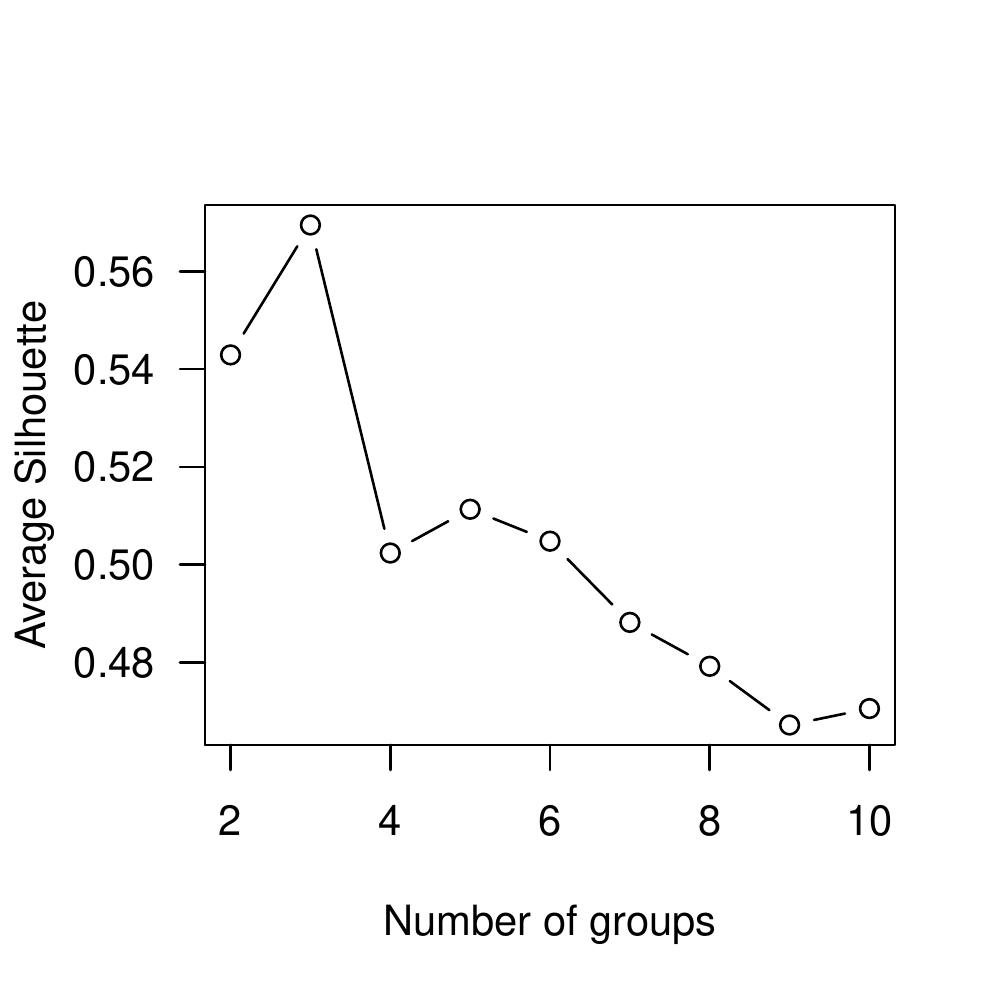}
\label{subfig:gz-averageSilhouette}
}
\vspace{-3pt}
\caption{Analysis of k-means clustering in the Galaxy Zoo project. Within-groups sum of squares and average Silhouette statistic as the number of groups (k) is varied.}
\label{fig:gz-clusterEvaluation}
\vspace{-5pt}
\end{figure}

The result of the quality of the clustering when the number of clusters varies between $2$ and $10$ is shown in Figure~\ref{fig:mw-clusterEvaluation} for The Milky Way Project and in Figure~\ref{fig:gz-clusterEvaluation} for Galaxy Zoo. These figures show that $5$ is the number of groups that best optimise the trade-off between the number of groups and the within-group sum of squares (Fig~\ref{subfig:mw-withinGroupsSumOfSquares} and~\ref{subfig:gz-withinGroupsSumOfSquares}). This number of groups also yields an Averaged Silhouette statistic of $0.53$ in The Milky Way Project (Fig.\ref{subfig:mw-averageSilhouette}) and $0.51$ in the Galaxy Zoo project (Fig.~\ref{subfig:gz-averageSilhouette}). These values indicate that a reasonable clustering structure has been found for both projects.


\subsection{Profiles}

\begin{figure}[htb]
\centering
\subfigure[The Milky Way Project]{
\includegraphics[width=0.48\linewidth]{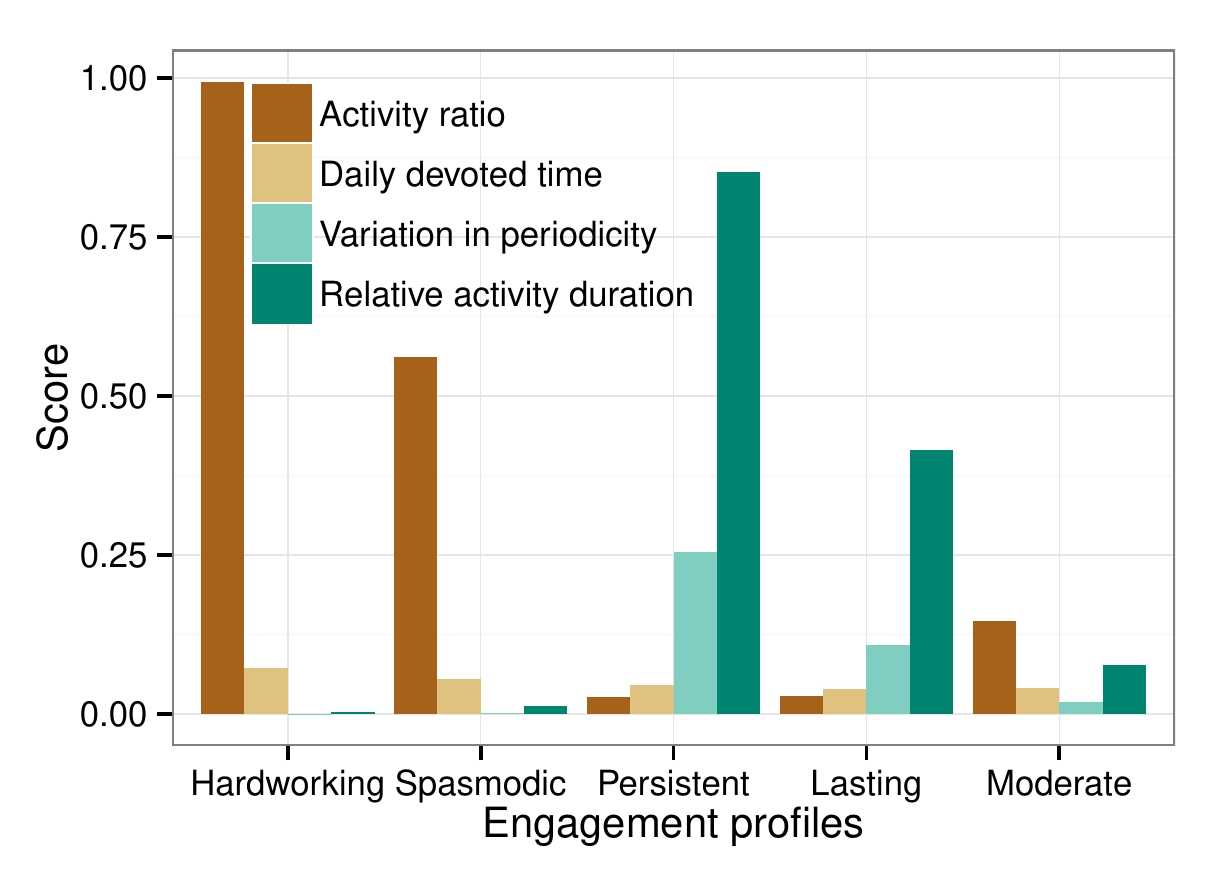}
\label{subfig:mw-centroids}
}
\subfigure[Galaxy Zoo]{
\includegraphics[width=0.48\linewidth]{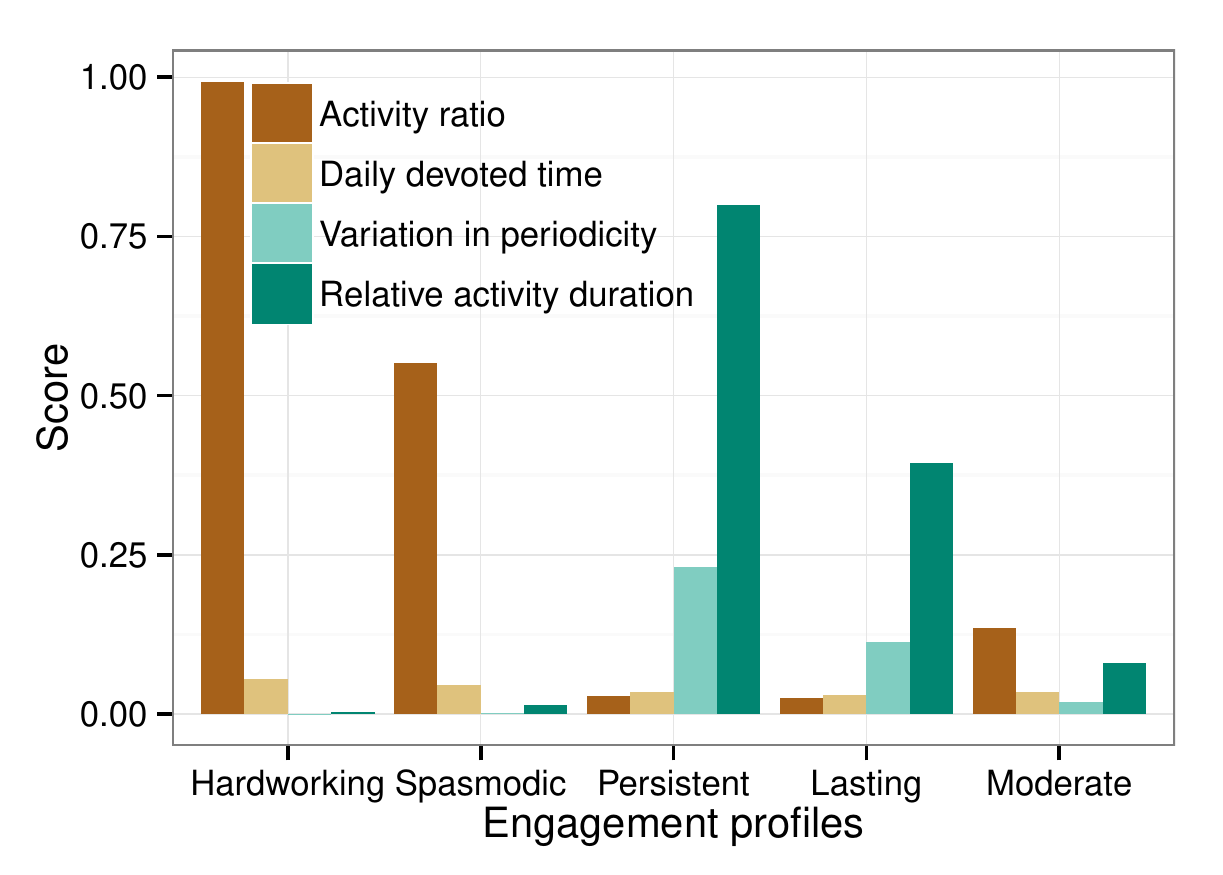}
\label{subfig:gz-centroids}
}
\caption{Score of each engagement profile in each engagement metric. Engagement profiles are represented by the centroids of groups of volunteers identified by the k-means algorithm in (a) The Milky Way Project and (b) Galaxy Zoo project.}
\label{fig:profiles}
\end{figure}

In order to understand the different groups uncovered by the clustering algorithm, we analyse: (i) the centroids that represent the groups; (ii) the correlation between each pair of volunteer engagement metrics for each group; and (iii) how the groups differ in terms of the number of volunteers and aggregate contribution. In this analysis, we established labels to the groups in order to put into pespective their main engagement characteristics. Thus, the groups represent different engagement profiles labelled as follows: hardworking engagement; spasmodic engagement, persistent engagement; lasting engagement; and moderate engagement. The general characteristics of these profiles are shown in Figure~\ref{fig:profiles}, Table~\ref{tab:correlations} and Table~\ref{tab:importance}.

Figure~\ref{fig:profiles} shows the centroids that represent each profile and how they differ in terms of engagement metrics. In each image, the horizontal axis stands for the engagement profiles, each bar representing one engagement metric, and the vertical axis indicates how the profiles score in the particular engagement metrics. Table~\ref{tab:correlations}, in turn, shows how the profiles differ in terms of correlation between their engagement metrics. Finally, Table~\ref{tab:importance} shows how the profiles differ in terms of the number of volunteers and how their aggregate contributions differ in terms of total working time devoted to the project. In the following paragraphs, we elaborate on these results by analysing each engagement profile in turn.

\begin{table}[htb]
\footnotesize
\caption{Spearman $\rho$ correlation between each pair of engagement metrics of volunteers within each engagement profile}
\label{tab:correlations}
\centering
\begin{tabular}{l|r| r | r | r |r }
\multicolumn{6}{c}{\bf The Milky Way Project}\\ \hline
\multirow{2}{*}{\bf Pair} &
\multicolumn{1}{c}{\bf Hardworking} &
\multicolumn{1}{|c}{\bf Spasmodic} &
\multicolumn{1}{|c}{\bf Persistent}&
\multicolumn{1}{|c}{\bf Lasting}&
\multicolumn{1}{|c}{\bf Moderate} \\
& \multicolumn{1}{c}{\bf$N = 1,535$} &
\multicolumn{1}{|c}{\bf$N = 1,060$}  &
\multicolumn{1}{|c}{\bf$N = 817$}  &
\multicolumn{1}{|c}{\bf$N = 844$}  &
\multicolumn{1}{|c}{\bf$N = 1,837$}\\ \hline \hline
$\rho(a, r)$ & -0.24*    & -0.38*      &-0.14*       & -0.26* & {\bf -0.74}*\\
$\rho(a, v)$ & {\bf -0.99}* & -0.22*      & 0.06\blank{5pt} &  0.39* & -0.13*\\
$\rho(a, d)$ & -0.07*    & -0.05\blank{5pt} & {\bf 0.43}* &  0.37* & 0.14*\\
$\rho(r, v)$ &  0.24*    & {\bf 0.59}* &-0.13*       & -0.04\blank{5pt}& {\bf 0.44}*\\
$\rho(r, d)$ &  0.14*    &  0.23*      &-0.09*       &  0.02\blank{5pt}& 0.01\blank{5pt}\\
$\rho(v, d)$ &  0.07*    &  0.29*      & 0.19*       &  0.31* & 0.21*\\\hline
\multicolumn{6}{c}{\bf }\\
\multicolumn{6}{c}{\bf Galaxy Zoo}\\
\hline
\multirow{2}{*}{\bf Pair} &
\multicolumn{1}{c}{\bf Hardworking} &
\multicolumn{1}{|c}{\bf Spasmodic} &
\multicolumn{1}{|c}{\bf Persistent}&
\multicolumn{1}{|c}{\bf Lasting}&
\multicolumn{1}{|c}{\bf Moderate} \\
& \multicolumn{1}{c}{\bf$N = 4,572$} &
\multicolumn{1}{|c}{\bf$N = 3,611$}  &
\multicolumn{1}{|c}{\bf$N = 3,783$}  &
\multicolumn{1}{|c}{\bf$N = 4,250$}  &
\multicolumn{1}{|c}{\bf$N = 7,331$}\\ \hline \hline
$\rho(a, r)$ & -0.30*       & -{\bf 0.45}* & 0.15*       & -0.23* &   {\bf -0.76}* \\
$\rho(a, v)$ & {\bf -0.99}* & -0.31*       & -0.26\blank{5pt} &  0.27* &   -0.12*\\
$\rho(a, d)$ & -0.10*       &  0.03\blank{5pt} &  0.33*      &  0.30* &   0.19*\\
$\rho(r, v)$ &  0.30*       &  {\bf0.66}*  & -0.12*      &  0.00\blank{5pt}&   {\bf0.43}*\\
$\rho(r, d)$ &  0.07*       &  0.17*       & 0.08*       &  0.02\blank{5pt}&   -0.05*\\
$\rho(v, d)$ &  0.10*       &  0.26*       & -0.01\blank{5pt}& 0.16* &   0.16*\\\hline
\end{tabular}
\begin{tablenotes}
\centering
\item {\it Note 1:} *Spearman' $\rho$ significant coefficient of correlation (p-value $< 0.05$).
\item {\it Note 2:} Moderate and strong correlations are highlighted in boldface.
\end{tablenotes}
\end{table}

\begin{table}
\footnotesize
\caption{Profiles importance in terms of the number of volunteers and their devoted time}
\label{tab:importance}
\centering
\begin{tabular}{l | r | r | r | r }
            \hline
            \multirow{2}{*}{\bf Profiles}& \multicolumn{2}{c}{\bf The Milky Way  Project} & \multicolumn{2}{|c}{\bf Galaxy Zoo} \\ \cline{2-5}
            & \multicolumn{1}{c|}{\bf \#Volunteers} & \multicolumn{1}{c|}{\bf Devoted time} & \multicolumn{1}{c|}{\bf \#Volunteers} & \multicolumn{1}{c}{\bf Devoted time}\\ \hline
Hardworking & 1,535 (25.19\%)    & 2,030.26 (13.86\%)
            & 4,572 (19.42\%)    & 4,857.49\blank{8pt}(9.44\%)\\
Spasmodic   & 1,060 (17.40\%)    & 1,912.05 (13.05\%)
            & 3,611 (15.34\%)     & 6,061.40 (11.78\%)\\
Persistent  & 817 (13.41\%)      & {\bf 5,846.58 (39.91\%)}
            & 3,783 (16.07\%)    & {\bf 23,757.64 (46.16\%)}\\
Lasting     & 844 (13.85\%)      &  2,273.10 (15.52\%)
            & 4,250 (18.05\%)    & 8,168.95 (15.87\%)\\
Moderate    &{\bf1,837 (30.15\%)}& 2,588.28 (17.67\%)
            &{\bf 7,331  (31.13\%)} & 8,621.64  (16.75\%)\\ \hline

\multicolumn{1}{r|}{\it sum} & 6,093 \blank{4pt}(100\%)\blank{4pt} &  14,650.27 \blank{4pt}(100\%)\blank{4pt} & 23,547 \blank{4pt}(100\%)\blank{4pt}&  51,467.12 \blank{4pt}(100\%)\blank{4pt}  \\\hline
\end{tabular}
\begin{tablenotes}
\centering
\item {\it Note:} The highest number of volunteers and the longest devoted time for each project are highlighted in boldface.
\end{tablenotes}
\end{table}

{\bf Hardworking} engagement. Volunteers who exhibit a hardworking engagement profile have larger activity ratio and shorter relative activity duration compared to others profiles (Fig~\ref{fig:profiles}). Such metrics indicate that volunteers in this profile work hard when they come into the project, but may leave the project soon. This engagement profile also exhibits low variation in periodicity. This means that volunteers who exhibit this engagement profile return to the project to perform more tasks in nearly equal intervals of time, which makes the time of return of these volunteers fairly predictable. Other intrinsic feature of this group of volunteers is a very strong negative correlation between activity ratio and variation in periodicity ($\rho(a,v) = -0.99$, in both projects). This correlation indicates that the more days the volunteers return to the project to perform tasks, the less variable are the time intervals between their active days.

{\bf Spasmodic} engagement. This engagement profile is distinguished by a relatively high activity ratio and low activity duration (Fig~\ref{fig:profiles}). This group of volunteers exhibits a positive correlation between relative activity duration and variation in periodicity. This correlation is moderate ($\rho(r,v) = 0.59$) in the Milky Way Project and strong ($\rho(r,v) = 0.66$) in the Galaxy Zoo project (Table~\ref{tab:correlations}). These correlations indicate that the longer the period of time the volunteers remain linked to the project, the more erratic is the periodicity of their return to the project within this period. All these characteristics indicate that contributions of volunteers exhibiting this profile typically takes place during a short period of time and with irregular periodicity within this period.

{\bf Persistent} engagement. Persistent engagement is characterised by the largest relative activity duration, the highest variation in period, and a short activity ratio (Fig~\ref{fig:profiles}). Thus, volunteers with a persistent engagement profile remain linked to the project for a long interval of time but are active only a few days within this interval. Considering these engagement metrics, persistent engagement may be seen as the opposite of hardworking engagement. In both projects, a small percentage of all the volunteers fall in this engagement profile: $13.41\%$ in The Milky Way Project and $16.07\%$ in the Galaxy Zoo project. Together, these volunteer stands for the largest percentage of the total working time devoted to each project, $39.91\%$ in The Milky Way Project and $46.16\%$ in the Galaxy Zoo project (Table~\ref{tab:importance}). It is the most important profile in terms of devoted working time.

{\bf Lasting} engagement. This is the engagement profile of volunteers exhibiting comparatively high relative activity duration and variation in periodicity (Fig~\ref{fig:profiles}). This kind of volunteers show an activity ratio similar to that exhibited by the volunteers who stay longer in the project (persistent engagement) but remain in the project during a shorter period of time. Finally, this is the only engagement profile showing very weak or weak correlation between all pairs of metrics in both projects (Table~\ref{tab:correlations}).

{\bf Moderate} engagement. As shown in Figure~\ref{fig:profiles}, this engagement profile has no particularly distinguishable engagement metrics. Compared to the other profiles, moderate volunteers exhibit intermediate values in all engagement metrics. One important characteristic of moderate engagement is a strong negative correlation between activity ratio and relative activity duration. This correlation is $\rho(a,r) = -0.74$ in The Milky Way Project and $\rho(a,r) = -0.76$ in Galaxy Zoo (Table~\ref{tab:correlations}). These values indicate that the degree of volunteer engagement in this profile falls with increased engagement duration. Hence, the more days the volunteers return to the project to perform tasks, the shorter is the total period of time that they remain linked to the project. This engagement profile is exhibited by most volunteers in both studied projects: nearly $30\%$ of the volunteers in The Milky Way Project and $31\%$ in Galaxy Zoo fall into this engagement profile (Table~\ref{tab:importance}).

\subsection{Discussion}

Our results show that volunteers in the studied projects share several similarities and differences in terms of engagement. The identified profiles of engagement put into perspective such similarities and differences. Furthermore, they help us to better understand how the different engagement patterns result in different levels of aggregated contribution to the projects. Several practical and research discussions can be done from this analysis. We focus on four of them, which are: profile-oriented volunteers' recruitment, personalised engagement strategies, psychological factors behind the engagement profiles, and external validity of the results.

{\bf Profile-oriented volunteers' recruitment}. It is natural that scientists running citizen science projects that require human computation want to devote more effort in recruiting volunteers who exhibits a desired engagement profile. It is still the most important aspect when they want to optimise the tradeoff between the costs of recruiting volunteers and the benefit of having all tasks of the project performed as soon as possible~\cite{Ponciano:Jisa:2014}. Studies have been devoted to understanding how different disclosure campaigns (e.g. traditional media and online media~\cite{Robson:2013}) differ in terms of the type of volunteers they attract. In a similar direction, it is also important to know how different disclosure campaigns differ in terms of the engagement profile of the volunteers they attract. For example, could a disclosure campaign based on sending e-mails to people interested in the theme of the project (e.g., astronomy, biology) attract more persistent volunteers than advertising campaigns in traditional media? Other important aspects that can be taken into account in optimising volunteer recruitment is human homophily~\cite{McPherson:2001}, which is the principle that humans tend to be similar to their friends in several aspects. Perhaps taking homophily into account one could motivate volunteers with a desired engagement profile to recruit volunteers among his/her relatives, friends, and colleagues with a similar profile? Hence, new and more effective recruitment procedures might be brought forth with an increased knowledge on volunteer engagement profiles.

{\bf Personalised engagement strategies.} Besides recruiting more suitable volunteers, it is also important to keep existing volunteers engaged. The impact of management practices on volunteer engagement is a widely discussed issue in volunteerism literature~\cite{Clary:1992,cravens:2000}. Such practices are implemented by volunteer supervisors in a way that takes into account the specific behaviour of each volunteer, aiming thereby at enriching the volunteer experience and satisfying organizational needs. By showing that volunteers in human computation for citizen science projects behave very differently from each other, this study encourages the development of a component to manage the engagement of volunteers in such projects. This component would incorporate personalised engagement strategies~\cite{Fischer2001,lopez:2012,mao:2013} derived from the volunteer engagement profiles uncovered in the present work. The component could also both monitor the contribution behaviour of each volunteer and, when necessary, automatically trigger a suitable engagement strategy. Prospective volunteers with different behaviour profiles should be approached with different engagement strategies, which could focus on e.g. encouraging a reduction or an improvement of their engagement.

Strategies can focus on encouraging a reduction of volunteer engagement when, for example, some volunteers start to compromise too much of their time to the project, which could perhaps have a negative impact on the rest of his/her social life, in the worst case leading to a state of burnout~\cite{gonzalez:2006,simpson:2009}. Fortunately, this is not the typical situation in the two projects we have studied; even volunteers with a hardworking engagement profile devote typically less than $21$ minutes per day to the project, which is not alarming. It is important that this kind of behavior can be monitored, and, if necessary, strategies are put in place to deal with the potential harm that this can bring to volunteers. When volunteers exhibit a suitable engagement profile, it is very important to recognize their contributions in order to keep them engaged~\cite{wilson:2000,rotman:2012}. Strategies can also focus on encouraging the improvement of volunteer engagement when volunteers exhibit a level of engagement below project average. This occurred frequently in the projects we have studied. Each volunteer engagement profile shows a lower level of engagement than the moderate engagement profile in at least one engagement metric.

There is a large body of work on strategies for encouraging contribution to online projects. Many of those strategies are discussed by~\citet{kraut:2012}. Example of strategies are (i) sending a message to the volunteers asking them for more contribution; or (ii) providing volunteers online in the project with specific and highly challenging goals, e.g. executing a number $n$ of tasks before logoff. One non trivial question that must be answered before putting a strategy to work is which engagement metrics one wishes to improve. Discovering the engagement profiles of the volunteers enables finding out in which engagement metric each profile falls short, and to decide which strategy to develop focusing on each volunteer profile. The correlations between the engagement metrics in each engagement profile tell us how other engagement metrics are affected when strategies are put into practice to improve one specific engagement metric. They also allow one to assess, for example, the additional gains that could be obtained from the multiplicative effects resulting from relationships between various metrics.

{\bf Psychological factors behind the engagement profiles.} As we discussed early, some studies have sought to understand the motivation of volunteers to participate in human computation for citizen science projects~\cite{raddick:2010,rotman:2012,jennett:2014}. Our results open a new perspective for such studies. Given that we have shown that volunteers exhibit different engagement profiles, new studies on the motivation factors can be conducted considering the engagement peculiarities of each profile. One major question to be answered in such studies is which motivations may lay behind each engagement profile. This calls for a more theoretical perspective, for example: (i) considering self-determination theory~\cite{Deci2000}, are persistent volunteers more extrinsically motivated than the volunteers who exhibit other engagement profiles? or (ii) considering self-efficacy theory~\cite{bandura1977}, why do hardworking volunteers expend much effort in the short term, but do not sustain their engagement in the long term. Besides complementing our understanding of volunteer engagement, such studies may provide information about volunteer motivation and experience in the projects.

In the profiles' analysis, we observe an opposition between degree of engagement and duration of engagement. Such opposition is clear in two main points: $1)$ very strong negative correlation between activity ratio and activity duration in the moderate engagement profile; $2)$ the opposition between the characteristics of hardworking engagement and persistent engagement. The negative correlation between activity ratio and activity duration in the moderate engagement profile indicates that participating in the project with a high frequency rate and remaining a long time in the project are contradictory characteristics. It can also be observed in the opposition between hardworking volunteers and persistent volunteers. Hardworking volunteers show a higher degree of engagement, but with a shorter duration. Persistent volunteers, on the contrary, show a lower degree of engagement but during a longer time period. It is important to understand the factors behind this opposition and to ask if there are situations in which the volunteers would present both a high degree and a long duration of engagement.

{\bf External validity.} Here we discuss about the generality of our study considering two main aspects: (i) whether the methodology we have proposed to measure the engagement of volunteers and identify their engagement profiles can be applied in other projects; and (ii) whether the results obtained in the case study with data collected from Galaxy Zoo and The Milky Way Project can be generalised to other human computation for citizen science projects.

The methodology we have proposed is based on theoretical frameworks that support the study of voluntarism~\cite{clary:1998,wilson:2000} and  human engagement~\cite{bandura1977,Brien:2008}. We draw on such frameworks to derive metrics for measuring the engagement of volunteers and to uncover engagement profiles from grouping them. In the case study conducted with data collected from Galaxy Zoo and The Milky Way Project, this methodology shown to be satisfactory in uncovering groups of volunteers that bring to light the main similarities and differences among them. Thus, studies seeking such quantitative analysis of the engagement can take advantage of this methodology.

Regarding the generality of the engagement profiles, there are two aspects that reinforce the idea that these types of profiles are more generic and thus can arise also in other types of projects. First, the same set of profiles have arisen in projects significantly different in terms of the tasks and the number of volunteers involved. Tasks in Galaxy Zoo are less time consuming than tasks in The Milky Way Project~\cite{Ponciano:CiSE:2014}. Galaxy Zoo has almost four times more volunteers than The Milky Way Project (Table~\ref{tab:descriptiveStatistics}), considering as volunteers those participants who have been active in at least two different days. As most of our results and conclusions are equivalent in both projects, the differences in the design of the tasks and in the number of volunteers have been shown not to affect the engagement profiles. Second, some profiles describe behaviours that are common in Web systems. For example, the observed fact that a small group of volunteers (persistent engagement) are responsible for the largest amount of contribution to the project has been shown to be valid also elsewhere~\cite{Eszter:2008,Trevor:2014}.

\section{Conclusions and Future Work}

In this study we answer three research questions: $1)$ how we can measure the level of engagement of volunteers during their interaction with a citizen science project that uses human computation;  $2)$ which different patterns of volunteer engagement behaviour can be identified and specified as typical volunteer profiles; and $3)$ how the identified volunteer engagement profiles can be exploited for designing strategies for increasing the engagement of volunteers in a project. We go through existing human engagement studies and, based on the concepts and theories put forward, we propose quantitative engagement metrics to measure different aspects of volunteer engagement, and use data mining algorithms to identify the different volunteer profiles in terms of the engagement metrics. We use this method to analyse the engagement of volunteers in two projects: Galaxy Zoo and The Milky Way Project. 

Our results show that volunteers in the studied projects share several similarities and differences in terms of engagement. We identify five distinct engagement profiles that put into perspective such similarities and differences. They are labelled as follows: hardworking, spasmodic, persistent, lasting, and moderate. These profiles differ among themselves according to a set of  metrics that we have defined for measuring the degree and duration of volunteer engagement. Regarding the distribution of the volunteers along the profiles, the highest percentage of volunteers falls into the moderate engagement profile, while only a few volunteers exhibit a persistent engagement profile. On the other hand, persistent volunteers account for the highest percentage of the total human effort dedicated to execute all the tasks in the project. Several discussions are drawn from our analysis, such as profile-oriented volunteers' recruitment, personalised engagement strategies, and psychological factors behind the engagement profiles.

Our analysis of volunteer engagement, based on log data, yielded a powerful framework for identifying the relevant patterns of volunteer engagement in human computation for citizen science projects. However, the current framework still presents some shortcomings that will be addressed in future work. We have focused on cognitive engagement of volunteers executing human computation tasks, but it is known that volunteers also contribute by creating additional content such as posts in project forums, which can be regarded as a form of social engagement. Assessing the behaviour of volunteers with regard to this type of engagement is also important. Finally, future work may be dedicated to analysing volunteer engagement in the context of other citizen science projects that use human computation. This analysis may give an answer to the question whether the set of engagement profiles we have identified on the basis of the two described projects is generic enough to be applied to the use of human computation for citizen science projects in general. Thus, we hope this study motivates further research on volunteer engagement in this type of projects.

\section{Acknowledgements}

We are indebted to Arfon Smith and Robert Simpson for providing us the dataset used in this study. We are also grateful to Herman Martins, Jussara Almeida, Nazareno Andrade, Jose Luis Vivas Frontana and the anonymous reviewers for their suggestions to improve several aspects of the manuscript. The authors would like to acknowledge the financial support received from CNPq/Brazil, CAPES/Brazil, and the European Union Seventh Framework Programme through the SOCIENTIZE project (contract RI-312902).

\bibliography{hcj-submission2-v2}
\end{document}